\documentstyle[12pt]{article}
\begin{document}
\rightline {\bf DFTUZ/95/13}
\vskip 2. truecm
\centerline{\bf THE GAUGED NAMBU-JONA LASINIO MODEL: A MEAN FIELD }
\centerline{\bf CALCULATION WITH NON MEAN FIELD EXPONENTS}
\vskip 2 truecm
\centerline { V.~Azcoiti$^a$, G. Di Carlo$^b$, A. Galante$^{c,b}$, 
A.F. Grillo$^d$, V. Laliena$^a$ and C.E. Piedrafita$^b$}
\vskip 1 truecm
\centerline {\it $^a$ Departamento de F\'\i sica Te\'orica, Facultad 
de Ciencias, Universidad de Zaragoza,}
\centerline {\it 50009 Zaragoza (Spain).}
\vskip 0.15 truecm
\centerline {\it $^b$ Istituto Nazionale di Fisica Nucleare, 
Laboratori Nazionali di Frascati,}
\centerline {\it P.O.B. 13 - Frascati 00044 (Italy). }
\vskip 0.15 truecm
\centerline {\it $^c$ Dipartimento di Fisica dell'Universit\'a 
dell'Aquila, L'Aquila 67100 (Italy)}
\vskip 0.15 truecm
\centerline {\it $^d$ Istituto Nazionale di Fisica Nucleare, 
Laboratori Nazionali del Gran Sasso,}
\centerline {\it Assergi (L'Aquila) 67010 (Italy). }
\vskip 3 truecm
\centerline {ABSTRACT}
We analyse the phase diagram of the lattice gauged Nambu-Jona Lasinio model 
with the help of a mean field approximation plus numerical simulations.
We find a phase transition line in the coupling parameters space separating
the chirally broken phase from the symmetric phase, which is in good 
qualitative agreement with results obtained in the quenched-ladder 
approximation. The mean field approximation relates the critical exponents 
along the continuous phase transition line with the mass dependence of the 
chiral condensate in the Coulomb phase of standard noncompact $QED$. Our 
numerical results for noncompact $QED$ strongly 
suggest non mean field exponents along the critical line.

\vfill\eject

\par
The gauged Nambu-Jona Lasinio $(GNJL)$ model has become increasingly 
interesting in recent time, one of the reasons for this increasing interest    
being the possibility to define a strongly coupled $QED$ with non trivial 
dynamics \cite{TODOS}. 
In fact if a non gaussian fixed point exists in non compact $QED$, 
the naive dimensional analysis does not applies. Therefore operators of 
dimension higher than four, which are non renormalizable in perturbation 
theory, could acquire anomalous dimensions and become renormalizable 
\cite{BARDEEN}. A good 
candidate is the four Fermi interaction, which when added to the standard 
$QED$ lagrangian preserving the continuous chiral symmetry, gives us the 
$GNJL$ model.

The lattice action for the $GNJL$ model with noncompact gauge fields and 
staggered fermions reads

$$S=\frac{\beta}{2}\sum_{n,\mu<\nu}\Theta_{\mu\nu}^2(n)\:+\:\bar\chi\Delta
(\theta)\chi\:+\:m\bar\chi\chi
\:-\:G\sum_{n,\mu}\bar\chi_n\chi_n\bar\chi_{n+\mu}\chi_{n+\mu}.
\eqno(1)$$

\noindent
where $\Theta_{\mu\nu}^2$ is the standard noncompact plaquette action, 
$\beta$ the inverse square coupling, $\Delta(\theta)$ the massless 
Dirac operator for 
Kogut-Susskind fermions and $G$ the four fermion coupling.

In the chiral limit, $m=0$, this action is invariant under the continuous 
transformations 

$$\chi_n\;\;\rightarrow\;\;\chi_n\: e^{i\alpha (-1)^{n_1+\ldots+n_d}}
\;\;\;\;\;\;\;
\;\;\bar\chi_n\;\;\rightarrow\;\;\bar\chi_n\: e^{i\alpha (-1)^{n_1+\ldots
+n_d}}
\eqno(2)$$

\noindent
which define a continuous chiral $U(1)$ symmetry group.

The vacuum expectation value of the chiral condensate is given by the 
following ratio of path integrals over the Grassmann and gauge fields

$$ \langle\,\bar\chi\chi\,\rangle
=\frac{\int\,[d\theta d\bar\chi d\chi]\,e^{-S}\,\frac{1}{V}\sum_n
\bar\chi_n\chi_n}{\int\,[d\theta d\bar\chi d\chi]\,e^{-S}}
\eqno(3)$$

The main technical difficulty when computing vacuum averages as (3) in the 
$GNJL$ model comes from the fact that the action (1) is not a bilinear of 
the fermion fields. The standard procedure consists in the introduction of 
an auxiliary vector field which allows to bilinearize the fermion action. 
The prize to pay for that is that we have one more field to include in the 
numerical simulations of this model that besides the number of free 
parameters 
$(\beta, m, G)$, makes it difficult to analyse this model with reasonable 
computer resources \cite{EDIN}.

Alternatively, we can perform a standard mean field approximation which 
also bilinearizes the action (1). Following the mean field 
technique, we make in (1) the following substitution

$$G\sum_{n,\mu}\bar\chi_n\chi_n\bar\chi_{n+\mu}\chi_{n+\mu}\;\;\rightarrow
\;\;2dG\langle\,\bar\chi\chi\,\rangle
\sum_n\bar\chi_n\chi_n
\eqno(4)$$

\noindent
where $d$ is the space-time dimension. The action (1) becomes in this way a 
bilinear in the fermion fields and the path integral over the Grassmann 
variables can be done by means of the Matthews-Salam formula.

The $v.e.v.$ of the chiral condensate (3) after the substitution 
of the mean field approximation (4) in the 
action (1) is given by

$$\langle\,\bar\chi\chi\,\rangle
=-\frac{\int\,[d\theta]\,e^{-\frac{\beta}{2}\sum\Theta_{\mu\nu}^2(n)}
\det[\Delta+(m-8G\langle\,\bar\chi\chi\,\rangle)
I]\frac{1}{V}tr\frac{1}{\Delta+(m-8G\langle\,\bar\chi\chi\,\rangle)I}}
{\int\,[d\theta]\,e^{-\frac{\beta}{2}\sum\Theta_{\mu\nu}^2(n)}
\det[\Delta+(m-8G\langle\,\bar\chi\chi\,\rangle)I]},
\eqno(5)$$

\noindent
which after simple algebraic operations can be written as

$$\langle\,\bar\chi\chi\,\rangle
=-2(m-8G\langle\,\bar\chi\chi\,\rangle)\,
\left\langle\,\frac{1}{V}\sum_{j=1}^{V/2}\frac{1}
{\lambda_j^2+(m-8G\langle\,\bar\chi\chi\,\rangle)^2}\,\right\rangle
\eqno(6)$$

\noindent
where the sum in (6) runs over all positive eigenvalues of the massless 
Dirac operator and the integration measure in the $v.e.v.$ includes the 
fermionic determinant of standard noncompact $QED$, evaluated at the 
effective mass $\bar{m}=m-8G\langle\,\bar\chi\chi\,\rangle$. 
In the chiral limit 
$m=0$, equation (6) becomes

$$\langle\,\bar\chi\chi\,\rangle=
16G\langle\,\bar\chi\chi\,\rangle\,
\left\langle\,\frac{1}{V}\sum_{j=1}^{V/2}\frac{1}
{\lambda_j^2+64G^2{\langle\,\bar\chi\chi\,\rangle}^2}\,\right\rangle
\eqno(7)$$
 
\vskip 1truecm
\leftline{\bf 1. The phase diagram}
\par
Equation (7) is always verified if $\langle\,\bar\chi\chi\,\rangle=0$, and 
this is the only solution in the symmetric phase. In the broken phase where 
$\langle\,\bar\chi\chi\,\rangle\neq 0$, the $v.e.v.$ of the chiral 
condensate 
will be given by the solution of the following equation

$$1=16G\,\left\langle\,\frac{1}{V}\sum_{j=1}^{V/2}\frac{1}
{\lambda_j^2+64G^2{\langle\,\bar\chi\chi\,\rangle}^2}\,
\right\rangle
\eqno(8)$$

\noindent
which gives for the critical line, where the chiral condensate vanishes 
continuously, the following expression

$$G_c(\beta)=\frac{1}{\frac{16}{V}\,\langle\,\sum_{j=1}^{V/2}\frac{1}
{\lambda_j^2}\,\rangle}
\eqno(9)$$

The existence of this critical line in the $GNJL$ model was discovered some 
time ago \cite{CRITICAL} in the continuum formulation using the 
quenched-ladder approximation.

Let us discuss qualitatively the phase diagram. For $\beta<\beta^{0}_c$, 
where $\beta^{0}_c$ is the critical coupling at $G=0$, the symmetry is 
always 
spontaneously broken since in this case equation (8) has a non vanishing 
solution for any $G\neq 0$. On the other side, the symmetry is also 
spontaneously broken in the $G\rightarrow\infty$ limit since in this limit 
equation (8) can be written as

$$1=16G\frac{1}{64G^2{\langle\,\bar\chi\chi\,\rangle}^2}\;
+\;O(\frac{1}{G^2})
\eqno(10)$$

\noindent
from which it follows that 

$$ \langle\,\bar\chi\chi\,\rangle\sim\frac{1}{2\sqrt{G}}
\eqno(11)$$

At $\beta = \infty$ the theory can be solved analytically in this 
approximation. The value of the critical four fermion coupling is in this 
limit $G_c= 0.2017$. For $G$ values smaller than this value, the symmetry 
is restored.

In Fig. 1 we present our numerical results for the phase diagram in the 
$\beta, G$ plane. The critical line has been obtained by computing 
numerically 
the $v.e.v.$ of the sum of the inverse square eigenvalues (eq. (9)), which 
is proportional to the chiral transverse susceptibility of the standard 
noncompact $QED$ in the chiral limit. The numerical simulations where 
performed using the $MFA$ approach \cite{MFA}, which 
allows to do computations in the chiral limit. We refer the interested 
reader to the extended bibliography on this subject \cite{MFA} 
and especially to the ref. \cite{SUSCEP} where the computation of the 
chiral 
susceptibility and the determination of the critical coupling in noncompact 
$QED$ is discussed in detail.

\vskip 1truecm
\leftline{\bf 2. The critical exponents}
\par
The phase diagram of Fig. 1 is in good qualitative agreement with the 
corresponding phase diagram obtained in the quenched-ladder 
approximation \cite{CRITICAL}. Using this analytical 
approach, a line of critical 
points with continuously varying critical exponents was found in 
\cite{HOMBRE}, 
the intersection point of this line with the $G=0$ axis corresponding to an 
essential singularity \cite{BARDEEN}.

Later on, numerical simulations of noncompact $QED$ disproved the essential 
singularity behavior \cite{TODOS}, putting in evidence the limitations 
of the 
quenched-ladder approximation. Since our approach contains weaker
approximations, we do hope to get more reliable results for the critical 
exponents.

In order to extract the critical exponents, we will start from the key 
equation of state (eq. (6)) relating the order parameter with the "external 
magnetic field" $m$ and the gauge and four fermion couplings. Using the 
previous notation we can write equation (6) as

$$\langle\,\bar\chi\chi\,\rangle=-2\bar{m}F(\beta,\bar{m})
\eqno(12)$$

\noindent
where the right hand side in (12) is just the chiral condensate in full 
noncompact $QED$ evaluated at the gauge coupling value $\beta$ and fermion 
mass $\bar{m}$. Concerning critical exponents the interesting physical 
region, as follows from the phase diagram of Fig. 1, is $\beta>\beta^{0}_c$ 
(Coulomb phase of noncompact $QED$).

Since we are interested in the critical region ($m\rightarrow 0$, 
$\langle\,\bar\chi\chi\,\rangle\rightarrow 0$), we will analyze the 
behavior 
of $F(\beta,\bar{m})$ in the $\bar{m}\rightarrow 0$ limit. In this limit 
we can write

$$F(\beta,\bar{m})=F(\beta,0) + B\bar{m}^\omega\:+\:\ldots
\eqno(13)$$

The second term in (13) possibly contains also logarithmic 
contributions and $F(\beta,0)$ is half the massless transverse 
susceptibility in noncompact $QED$. Therefore we can write

$$F(\beta,0)=\frac{1}{V}\langle\sum 1/\lambda_j^2\rangle_{\beta,\bar{m}=0}= 
{1/16G_{c}(\beta)}
\eqno(14)$$

\noindent
where $G_c(\beta)$ in (14) stands for a generic point of the critical line 
in Fig. 1. Equation (13), after the substitution of $\bar{m}$ by 
$m-8G\langle\,\bar\chi\chi\,\rangle$, implies the following behavior for 
the chiral condensate in the $m\rightarrow 0$ limit

$$\langle\,\bar\chi\chi\,\rangle\sim m^{\frac{1}{\omega+1}}
\eqno(15)$$

\noindent
and therefore the $\omega$ and $\delta$ exponents are related by the 
equation 

$$\delta = \omega + 1
\eqno(16)$$

A straightforward calculation allows to compute also the 
magnetic $\beta_m$ and susceptibility $\gamma$ exponents, the final result 
being

$$\beta_m=\frac{1}{\omega},\;\;\;\;\;\;\;\gamma = 1
\eqno(17)$$

The hyperscaling relation $\gamma=\beta_{m}(\delta-1)$ is verified, as 
follows from (17).

The determination of the critical exponents of the
order parameter in our mean field 
approach reduces therefore to the determination of the $\omega$ exponent 
which 
controls the mass dependence of the chiral condensate in the Coulomb phase 
of noncompact $QED$. In the $\beta\rightarrow\infty$ limit of noncompact 
$QED$, the theory is free and the chiral condensate can be analytically 
computed. The well known result in this case ($\omega=2$ plus logarithmic 
corrections) implies mean field exponents for the end point of the phase 
transition line, with the following behavior for 
$\langle\,\bar\chi\chi\,\rangle_{\beta=\infty,G=G^{\infty}_c}$

$$m\sim {\langle\,\bar\chi\chi\,\rangle}^{3} 
\log \langle\,\bar\chi\chi\,\rangle
\eqno(18)$$

In the general case, the chiral condensate in the Coulomb phase of 
noncompact $QED$ $(\langle\,\bar\chi\chi\,\rangle_{NCQED})$ can be 
parameterized as follows

$$\langle\,\bar\chi\chi\,\rangle_{NCQED} = A(\beta) m + B(\beta) m^
{\omega+1}\:+\:\ldots
\eqno(19)$$

The first contribution in (19) is linear in $m$, as follows from the fact 
that the massless transverse susceptibility is finite in the Coulomb phase 
of noncompact $QED$. The next contribution can possibly have logarithmic 
corrections, as happens in the $\beta\rightarrow\infty$ limit where it
becomes 
$m^{3}\log m$. In order to extract the $\omega$ exponent from the numerical 
simulations, we can use the results for the massless chiral transverse 
susceptibility \cite{SUSCEP} 
to fix $A(\beta)$ in (19) and fit the numerical results with 
eq. (19). This procedure has the inconvenient that higher order 
contributions 
in (19) can induce systematic errors in the determination of $\omega$. 
A better strategy is to measure the massless nonlinear susceptibility, 
defined 
as the third mass derivative of the chiral condensate. In this case we get 
only one contribution in the $m\rightarrow 0$ limit which is 
logarithmically 
divergent in the free field theory against a power divergence, which will 
appear if $\omega<2$.

Of course in a finite lattice, the non linear susceptibility is always 
finite. 
However simple finite size scaling arguments tell us that the nonlinear 
susceptibility should diverge logarithmically with the lattice size in the 
free field case whereas a power divergence with the lattice size is 
expected in the case $\omega<2$. In Fig. 2 we have plotted our results for 
the nonlinear susceptibility $\chi_{nl}$ of noncompact 
$QED$ against the lattice 
size at $\beta=0.237$, a value which is unambiguously in the Coulomb phase 
of this model \cite{TODOS}. 
This is a log-log plot and the four points correspond to 
lattice sizes 4, 6,8 and 10. As it is shown in the figure, the four points 
are very well fitted by a straight line, this implying that $\omega<2$. 

Due to the potentialities of the $MFA$ method, we have computed vacuum 
expectation values of other operators, which can be considered as 
generalizations of a term contributing to the massless nonlinear 
susceptibility. More precisely 
we have defined $\chi_q$ by the expression

$$\chi_{q} = \frac{1}{V}\left\langle\,\sum_{j=1}^{V/2}\frac{1}
{\lambda_j^q}\,\right\rangle
\eqno(20)$$

When $q=4$, we get one of the contributions to the standard massless 
nonlinear susceptibility. In the 
general case we can write this vacuum expectation value as an integral 
over the spectral density of eigenvalues in the following way

$$\frac{1}{V}\left\langle\,\sum_{j=1}^{V/2}\frac{1}
{\lambda_j^q}\,\right\rangle = \int \frac{\rho(\lambda)}{\lambda^q} 
d\lambda
\eqno(21)$$

\noindent
and if the density of eigenvalues $\rho(\lambda)$ behaves like $\lambda^p$ 
near the origin, $\chi_q$ will diverge when $q>p+1$. In such a case and 
for lattices of finite size, we expect for $\chi_q$ the following 
behaviour with the lattice size $L$

$$\chi_{q} \sim L^{\alpha(q-p-1)}
\eqno(22)$$

\noindent
where $\alpha$ in (22) is some positive number.

It is interesting to note that the $p$ exponent which controls the 
small $\lambda$ behavior of the spectral density $\rho(\lambda)$, can be 
related to the $\omega$ exponent by the following equations 

$$\omega = p-1 (p\leq 3)$$

$$\omega = 2 (p>3)
\eqno(23)$$

These relations allow to extract the $\omega$ exponent from the finite size 
behavior of the generalized nonlinear susceptibility $\chi_q$.

In Fig. 3 we have plotted our results for the inverse of the generalized 
nonlinear susceptibility $\chi_q$ against the inverse lattice size 
for $q$ values running from 2 to 4 and $\beta=0.237$. The solid lines in 
this figure correspond to a 
fit of all the points at any fixed $q$ with the function 

$$\chi_{q}^{-1}(L) = a_q+b_q L^{-c}
\eqno(24)$$ 

The results reported in this figure show that, 
in the infinite volume limit, 
the inverse generalized nonlinear susceptibility vanishes at large 
$q$ and 
is different from zero at small $q$, as expected. Fig. 4 is a plot of the 
extrapolated values of $\chi_q^{-1}$ (thermodynamical limit) against 
$q$. The 
critical value of $q$ at which $\chi_q^{-1}$ vanishes can be estimated from 
these results. Hence we get $q_c\sim 2.5$ at $\beta=0.237$, which implies 
$p\sim 1.5$ and $\omega\sim 0.5$. Using now the relations (16), (17) the 
following results for the order parameter critical exponents can be derived

$$\delta\sim 1.5,\;\;\;\;\;\;\;\beta_{m}\sim 2,\;\;\;\;\;\;\;\gamma=1, 
\eqno(25)$$

\noindent
values which are clearly outside the range of the mean field exponents. 

Non mean filed exponents within a mean field approximation might 
seem rather surprising at first sight. There is however no real 
contraddiction, since we have applied the mean field approximation to the 
fermion field, while fluctuations of the gauge field are 
fully taken into account 
in our numerical simulations. In the infinite $\beta$ limit, where the 
gauge field is frozen to the free field configuration, we get mean field 
exponents. However fluctuations of the gauge field at finite $\beta$ seem 
to play a fundamental role in driving critical exponents to non mean field 
values.

The picture which emerges from this calculation is that the critical 
exponents 
change continuously along the critical line of Fig. 1 from their mean 
field values (end point of the critical line) to some (non mean 
field) values 
at the critical point of noncompact $QED$. The $\delta$ exponent approaches 
its mean field value $(\delta=3)$ from below whereas the magnetic exponent 
approaches its mean field value $(\beta_{m}=0.5)$ from above \cite{SACHA}. 
Our results for several values of the gauge coupling $\beta$ 
suggest also that the value of $\delta$ increases systematically 
along the critical line with increasing $\beta$, in contrast with the 
magnetic 
exponent results which are systematically decreasing with $\beta$. 

In spite of the mean field approach for the fermion field, 
we believe that our qualitative picture is realistic.
It is in fact hard to imagine that 
non mean field exponents in a mean field approach will 
become mean field exponents after removing the mean field approximation, 
i.e. that restoring the full fluctuations of the fermion fields would 
drive back the critical exponents to mean field values.

A numerical analysis of the fermion-gauge-scalar model with compact 
$U(1)$ gauge symmetry done in 
\cite{JERSAK}, has shown the existence of a critical line  
separating a chirally broken phase from a symmetric phase. In the infinite 
gauge coupling limit the model is effectively described by the Nambu-Jona 
Lasinio model \cite{LEE} 
and therefore critical exponents at this point of the 
critical line are gaussian. However strong evidence for non mean field 
exponents has been found in \cite{JERSAK} near the tricritical point 
separating the second order line from the first order one. Contrary to 
our results for the Gauged Nambu-Jona Lasinio model, the critical 
behaviour along the critical line of the fermion-gauge-scalar model 
seems to be well described by the pure Nambu-Jona Lasinio model execpt 
near the tricritical point. However a slow variation of the critical 
exponents with incresing inverse gauge coupling $\beta$ in this model, 
like the one we have found in the Gauged Nambu-Jona Lasinio model, 
can not be excluded, we believe.

One important point which deserves further investigation in the $(GNJL)$ 
model is the physical 
origin of the non mean field behavior. In the $\beta\rightarrow\infty$ 
limit of noncompact $QED$ the second contribution to the chiral 
condensate behaves like $m^{3}\log m$ and this result will probably be true 
also in perturbation theory. Therefore some important role of gauge 
configurations topologically non equivalent to the free field configuration 
is suggested by our result $\omega<2$.

We thank CICYT (Spain) - INFN (Italy) collaboration for partial financial 
support to this work.
\vfill\eject

\vfill\eject
\vskip 1 truecm
\leftline{\bf Figure captions}
\vskip 1 truecm
{\bf Figure 1.} Phase diagram of the $GNJL$ model in the 
$\beta, G$ plane.

{\bf Figure 2.} Logarithm of the nonlinear susceptibility 
against the logarithm of the lattice size for lattice sizes 4,6,8, 10 and 
$\beta=0.237$.

{\bf Figure 3.} Inverse generalized nonlinear 
susceptibility against the inverse lattice volume at $\beta=0.237$.

{\bf Figure 4.} Infinite volume limit of the generalized 
nonlinear susceptibility against $q$ at $\beta=0.237$.

\end{document}